\documentclass[10pt,journal,compsoc]{IEEEtran}\usepackage{bm}

\usepackage{booktabs}
\usepackage{multirow}
\usepackage{enumitem}
\usepackage{bbm}
\usepackage{tabularx,booktabs}

\usepackage[section]{placeins}
\usepackage{float}

\usepackage{ifpdf}
 \ifpdf
 \else
 \fi

\ifCLASSOPTIONcompsoc
  \usepackage[nocompress]{cite}
\else
  \usepackage{cite}
\fi

\ifCLASSINFOpdf
   \usepackage[pdftex]{graphicx}
   
\else
\fi
\usepackage{amsmath}
\usepackage{amssymb}
\newenvironment{sequation}{\small\begin{equation}}{\end{equation}}

\usepackage{algorithmic}
\usepackage{algorithm}

\usepackage{array}

\ifCLASSOPTIONcompsoc
 \usepackage[caption=false,font=footnotesize,labelfont=sf,textfont=sf]{subfig}
\else
 \usepackage[caption=false,font=footnotesize]{subfig}
\fi
\usepackage{url}
\newcommand\MYhyperrefoptions{bookmarks=true,bookmarksnumbered=true,
pdfpagemode={UseOutlines},plainpages=false,pdfpagelabels=true,
colorlinks=true,linkcolor={black},citecolor={black},urlcolor={black},
pdftitle={Knowledge Graph-enhanced Sampling for Conversational Recommender System},
pdfsubject={Conversational Recommender System},
pdfauthor={Mengyuan Zhao, Xiaowen Huang*, Jitao Sang},
pdfkeywords={Conversational Recommender System, Knowledge Graph, Negative Sampling, Active Learning, Reinforcement Learning}}
\ifCLASSINFOpdf
\usepackage[\MYhyperrefoptions,pdftex]{hyperref}
\else
\usepackage[\MYhyperrefoptions,breaklinks=true,dvips]{hyperref}
\usepackage{breakurl}
\fi
\begin{document}
%
\title{Knowledge Graph-enhanced Sampling for Conversational Recommendation System}
%
%
%
%

\author{Mengyuan Zhao, 
Xiaowen Huang*, 
Lixi Zhu,
Jitao Sang,
Jian Yu
\IEEEcompsocitemizethanks{\IEEEcompsocthanksitem Mengyuan Zhao is with the School of Computer and Information Technology \& Beijing Key Lab of Traffic Data Analysis and Mining, Beijing Jiaotong University, Beijing 100044, China.\protect\\
E-mail: 19120453@bjtu.edu.cn.
\IEEEcompsocthanksitem Xiaowen Huang is with the School of Computer and Information Technology \& Beijing Key Lab of Traffic Data Analysis and Mining, Beijing Jiaotong University, Beijing 100044, China.\protect\\
E-mail: xwhuang@bjtu.edu.cn.
\IEEEcompsocthanksitem Lixi Zhu is with the School of Computer and Information Technology \& Beijing Key Lab of Traffic Data Analysis and Mining, Beijing Jiaotong University, Beijing 100044, China.\protect\\
E-mail: zlxxlz1026@gmail.com.
\IEEEcompsocthanksitem Jitao Sang is with the School of Computer and Information Technology \& Beijing Key Lab of Traffic Data Analysis and Mining, Beijing Jiaotong University, Beijing 100044, China.\protect\\
E-mail: jtsang@bjtu.edu.cn.
\IEEEcompsocthanksitem Jian Yu is with the School of Computer and Information Technology \& Beijing Key Lab of Traffic Data Analysis and Mining, Beijing Jiaotong University, Beijing 100044, China.\protect\\
E-mail: jianyu@bjtu.edu.cn.}
\thanks{Manuscript received July 1st, 2021; revised September 1st, 2021.\protect\\
(Corresponding author: Xiaowen Huang.)}
}

%
%

\markboth{IEEE TRANSACTIONS ON KNOWLEDGE AND DATA ENGINEERING, VOL. xx, NO. x, JULY 2021}%
{Zhao \MakeLowercase{\textit{et al.}}: Knowledge Graph-enhanced Sampling for Conversational Recommendation System}

%



\IEEEtitleabstractindextext{%
\begin{abstract}

The traditional recommendation systems mainly use offline user data to train offline models, and then recommend items for online users, thus suffering from the unreliable estimation of user preferences based on sparse and noisy historical data. Conversational Recommendation System (CRS) uses the interactive form of the dialogue systems to solve the intrinsic problems of traditional recommendation systems. However, due to the lack of contextual information modeling, the existing CRS models are unable to deal with the exploitation and exploration (E\&E) problem well, resulting in the heavy burden on users. To address the aforementioned issue, this work proposes a contextual information enhancement model tailored for CRS, called Knowledge Graph-enhanced Sampling (KGenSam). KGenSam integrates the dynamic graph of user interaction data with the external knowledge into one heterogeneous Knowledge Graph (KG) as the contextual information environment. Then, two samplers are designed to enhance knowledge by sampling fuzzy samples with high uncertainty for obtaining user preferences and reliable negative samples for updating recommender to achieve efficient acquisition of user preferences and model updating, and thus provide a powerful solution for CRS to deal with E\&E problem. Experimental results on two real-world datasets demonstrate the superiority of KGenSam with significant improvements over state-of-the-art methods.


\end{abstract}

}

\maketitle

\IEEEdisplaynontitleabstractindextext

%
\IEEEpeerreviewmaketitle

\renewcommand{\arraystretch}{1.3}


\ifCLASSOPTIONcompsoc
\IEEEraisesectionheading{\section{Introduction}\label{sec:introduction}}
\else
\section{Introduction}
\label{sec:introduction}
\fi

\IEEEPARstart{R}{ecommender} systems are the information filtering systems that help users filter a large number of invalid information to obtain information or items by estimating their interests and preferences. The mainstream traditional recommendation systems mainly use offline and historical user data to continuously train and optimize offline models, and then recommend items for online users. There are three main problems: the unrealistic estimation of user preferences based on sparse and noisy historical data\cite{2020WangDenoising,2020WangClicknotLike,2019LeeMeLU}, the ignorance of online contextual factors that affect user behavior\cite{2020CenControllableMultiInterest,2019GaoBLOMA,2019JagermanOffPolicyEvaluation}, and the unreliable assumption that users are aware of their preferences by default\cite{2013WangContingencyApproachInteractionModes}. Inspired by the dialogue systems, which focus on the user's real-time feedback data and obtains the user's current interaction intentions, a new recommendation method, Conversational Recommendation System (CRS), is trying to combine the interactive form of the dialogue systems with the recommendation task and becomes an effective means to solve the traditional recommendation problem. Through online interactive methods, CRS can guide and capture user' current preferences and interests by asking user whether he/she likes an item attribute or not, and utilize the user’s preferred attributes or rejected attributes to provide recommendation results that most fit users' current preferences in time. 

The core research difficulty of CRS is exploitation and exploration (E\&E) problem. Specifically, at each turn of the conversational session, CRS may choose to utilize the user preference information obtained during the previous interaction to output the current best recommended items, which is the exploitation of E\&E; or choose to continue to interact with users to obtain more real-time user preference information, which is the exploration of E\&E. The difficulty of solving CRS's E\&E problem lies in achieving a special Nash equilibrium in the field of recommendation systems: if CRS is too inclined to choose exploitation action ($recommend$) without understanding the current user preferences, there will be a lot of failed recommendations which poor user experience; in contrast, if CRS adopts excessive exploration action ($ask$), the interaction burden on users will be increased and the risk of user churn will become serious. At present, CRS researchers mostly use Deep Reinforcement Learning (DRL) to learn a conversational recommendation policy to solve the E\&E problem. Because DRL has the ability of sequential decision-making, and it naturally fits with the interaction form of CRS. From the experimental results of the previous work, the CRS based on DRL achieves much better results than other methods, and the state-of-the-art CRS methods are all based on DRL. However, the existing works of conversational recommendation fail to put forward an effective solution to E\&E problem and build an efficient and accurate CRS, due to the lack of information utilization of the interactive environment.

In this paper, we believe that sufficient modeling of environment information is the key point to solving the E\&E problem. Therefore, our goal is to build a CRS that can make good use of the information of contextual environment at each turn of conversational recommendations. From the perspective of data enhancement, we propose a novel solution named Knowledge Graph-enhanced Sampling (KGenSam), which solves the insufficient information modeling problem in CRS. In KGenSam, we introduce Knowledge Graph (KG) into the basic CRS framework as its interactive environment, and design two sampler modules to model and enhance the information of KG environment. Hence, CRS will learn to make E\&E decisions based on enhanced knowledge.

KGenSam adopts the sampling method of KG nodes to enhance knowledge in CRS. The design of the two samplers in KGenSam is inspired by solving the following two specific problems which prevent CRS from making the optimal decisions of E\&E:

\textbf{(1) What attributes should CRS ask to help the recommender fit the current user's preferences efficiently?}\cite{lei2020estimation}

In a conversational recommendation session, CRS needs to obtain sufficient user preference information as much as possible to achieve a successful recommendation that fits the current user preference. Obviously, on the premise of successful recommendation in the final round, the fewer attributes the CRS asks, the shorter the interaction rounds, and the higher the efficiency of CRS. We believe that every asking opportunity of CRS is extremely valuable, and CRS needs to obtain as much user preference information as possible at every $ask$ opportunity. In the design of this work, we naturally refer to the core idea of Active Learning, that is, when CRS chooses $ask$ action,  high-uncertainty attribute candidates that can not be judged by the current model need to be given priority attention. Because these attribute samples with high uncertainty will greatly improve the recommendation confidence after being annotated by users and will make great contributions to achieving the goal of obtaining sufficient user preferences with as little user interaction burden as possible. Therefore, to screen out the fuzzy samples with high uncertainty, we design an Active Sampler, which takes the sample distribution state in KG and conversation state as input and learns the features of fuzzy samples via active reinforcement learning.

\textbf{(2) How can CRS update the recommender precisely to fit the implicit preference information in user feedback?}

To train an accurate personalized recommender, both positive and negative samples of user feedback are required\cite{2012RendleBPR,2017HeFastMF}.The user interaction form of CRS naturally solves the one-class problem\cite{2008PanOneClass} of traditional recommendation that most user data is positive feedback.  In CRS, besides the user's positive feedback data, the user's explicit negative feedback data can also be directly obtained.  But CRS also presents a new research challenge. The number of conversation turns is often limited to more than a dozen, which causes the amount of user feedback data in a conversation recommendation session is also very limited. Generally, sparse user feedback and huge candidate item space require a lot of training data to train an effective recommender. How to update the recommender by using rare online feedback data has become a new challenge for researchers. Inspired by the idea of solving one-class problems by negative sampling in traditional recommendation systems, we try to supplement sparse interactive data by enhancing the positive and negative sample pairs in KG. We design a Negative Sampler, which refers to the hard negative samples mining method in image recognition. The Negative Sampler takes the sample distribution state in KG and conversation state as input, learns the features of hard negative samples via reinforcement learning, and constructs the positive and negative sample pairs enhanced from the user feedback data to accurately update the recommender. 

To evaluate the effectiveness of KGenSam, we conducted experiments on the Yelp and LastFM datasets. The experimental results show that KGenSam significantly outperforms the state-of-the-art CRS methods, indicating that KGenSam has the ability to enhance the user experience of the whole CRS. We release the codes, datasets and demo of the KGenSam-based online CRS at \href{https://github.com/ORZisemoji/KGenSam}{https://github.com/ORZisemoji/KGenSam}.

In a nutshell, this work makes the following main contributions:
\begin{itemize}
    \item We propose the KGenSam model, which provides a new angle for building CRS. By sampling in KG, KGenSam enhances the contextual knowledge of CRS and solves the E\&E problem caused by insufficient information modeling of CRS. To the best of our knowledge, it is the first time to introduce the knowledge graph-enhanced sampling method into the conversational recommendation.
    \item We introduce the active learning and the hard-negative sample mining methods into CRS. By incorporating the fuzzy samples and hard-negative samples in the CRS task, we facilitate the utilization of KG for CRS, which contributes to mine the user preference information in the KG environment to solve the data sparsity problem of the user feedback.
    \item Extensive experiments have been conducted on two benchmark datasets, demonstrating that KGenSam-based CRS is able to achieve better performance than state-of-the-arts with much fewer conversation turns, which indicates high sample efficiency.
    \item We implemented an online CRS to verify the effectiveness of our method in real application scenarios.
\end{itemize}

\section{Related work}

Recommender systems are the important applications of artificial intelligence, which can help users find the information that fits their preferences in the huge amount of data when there is no clear demand or the environment information is overloaded. 
At present, the mainstream recommendation systems are based on offline, historical user data, constantly optimize the offline performance, train the model, and then recommend items for online users. This offline update mode makes the recommender systems have inherent weakness \cite{2020WangDenoising,2020WangClicknotLike,2019LeeMeLU}, that is, it is difficult to fit the user's online behavior, and the user's real-time feedback will not be received in time\cite{2020CenControllableMultiInterest,2019GaoBLOMA,2019JagermanOffPolicyEvaluation,2013WangContingencyApproachInteractionModes}, so it is difficult for the systems to know the user's current intention and interest when using the recommender systems.

The development of dialogue technology in dialogue systems provides a new solution to the problems existing in traditional recommendation systems, which is called conversational recommendation. Conversational recommendation breaks the barrier of information asymmetry between systems and user in static recommendation systems through rich interaction behavior, and allows recommendation systems to capture user preferences dynamically in the interactive conversation with users. Based on the idea of conversational recommendation, CRS is a task-oriented and multi-rounds recommendation system, on the one hand, it explores users' current interest preferences, guides users to find new interest points, which realizes users' long-term retention; on the other hand, it also can receive feedback from users and update the recommender in real-time, which realizes dynamic learning of preferences. 

Recently years, researchers have proposed different CRS models designed from different perspectives. To define a standard form for conversation recommendation, Zhang et al.\cite{zhang2018towards} proposed a conversation search engine called SAUR. To solve the problem of user cold-start in the movie recommendation scenario, Li et al.\cite{li2018towards} designed a 
CRS that can interact with users through natural language. To obtain both long-term and real-time user preferences, Christakopoulou et al.\cite{christakopoulou2018q} designed a CRS that can take user interaction history as part of input data. After that, Sun et al.\cite{sun2018conversational} considered multi-rounds interaction with the user and designed a CRS that can obtain user attribute-level preferences multiple times. Lei et al.\cite{lei2020estimation} proposed a multi-rounds CRS framework named EAR, which can obtain both item-level and attribute-level feedback during the interaction. In order to take full advantage of conversational recommendation, Lei et al.\cite{lei2020interactive} introduced KG into CRS and proposed a new CRS framework called CPR. In order to make full use of user feedback information, Xu et al.\cite{xv2021FPAN} proposed FPAN framework, in which Gate Mechanism is used to dynamically modify the representations of
items and users based on the feedback information.

Overall, the research of CRS is committed to improving the satisfaction of online recommendations and the long-term retention rate of users in the system, and DRL has been widely used in CRS for learning conversational recommendation policy. However, the standard DRL model can not meet the needs of CRS, researchers are faced with the difficult convergence problem of CRS's training and low effectiveness problem of CRS's interaction. The main reason for these problems is the lack of clear rules in the contextual environment in CRS which leads to the lack of clear reward function and clear action space in the training of the reinforcement model. This can be summarized as the problem of insufficient environmental information modeling, which delays the updating of parameters in the training interaction, so CRS needs more interaction rounds to get user preference information and achieve a successful recommendation.





\section{Preliminaries}
\subsection{Problem Formulation}
In this subsection, We formally present interaction data and KG in our Task.

\textbf{Interaction Data.} $U$, $I$ and $A$ denote the set of users, items and attributes, respectively. Each item $i\in I$ is associated with a set of attributes $P_i\subseteq A$. 
For each user, we view each observed user-item interaction as a positive instance, while sampling a negative item to pair the user. And training interactive data is structured into two pairwise data sets, $O_I$ and $O_A$. $O_I$ and $O_A$ represent the user feedback on items and attributes, respectively:
\begin{equation}
    O_I = \{(u,i^+,i^-)|u\in U,i^+\in I^+,i^-\in I^-\}
\end{equation}

where $I^+$ denotes the sets of positive items, each of which has at least one historical interaction record with the user, and $I^-$ denote the sets of negative items, most of which are implicit negative feedback data of the user, and a few items are explicit negative feedback directly obtained by CRS.

\begin{equation}
    O_A = \{(u,p^+,p^-)|u\in U,p^+\in A^+,i^-\in A^-\}
\end{equation}

where $A^+$ denotes the sets of positive attributes that are the attributes of positive items or the user preferences $P_u$ acquired during the conversational recommendation session, and $A^-$ denote the sets of negative attributes that are the attributes of negative items or the  $ask$ attributes rejected by the user during the conversational recommendation session.

\textbf{Knowledge Graph.} As prior efforts\cite{2018ZhaoKB4Rec} show, items in user-item interaction data can be aligned with the corresponding entities in external KG. With such alignments, the external knowledge of item attributes as well as the interaction data are organized in the form of heterogeneous KG\cite{2019CaoUnifying}.

\subsection{Background settings of CRS in Our Task}
\begin{figure}[htb]
\centering
\includegraphics[width=3.7in]{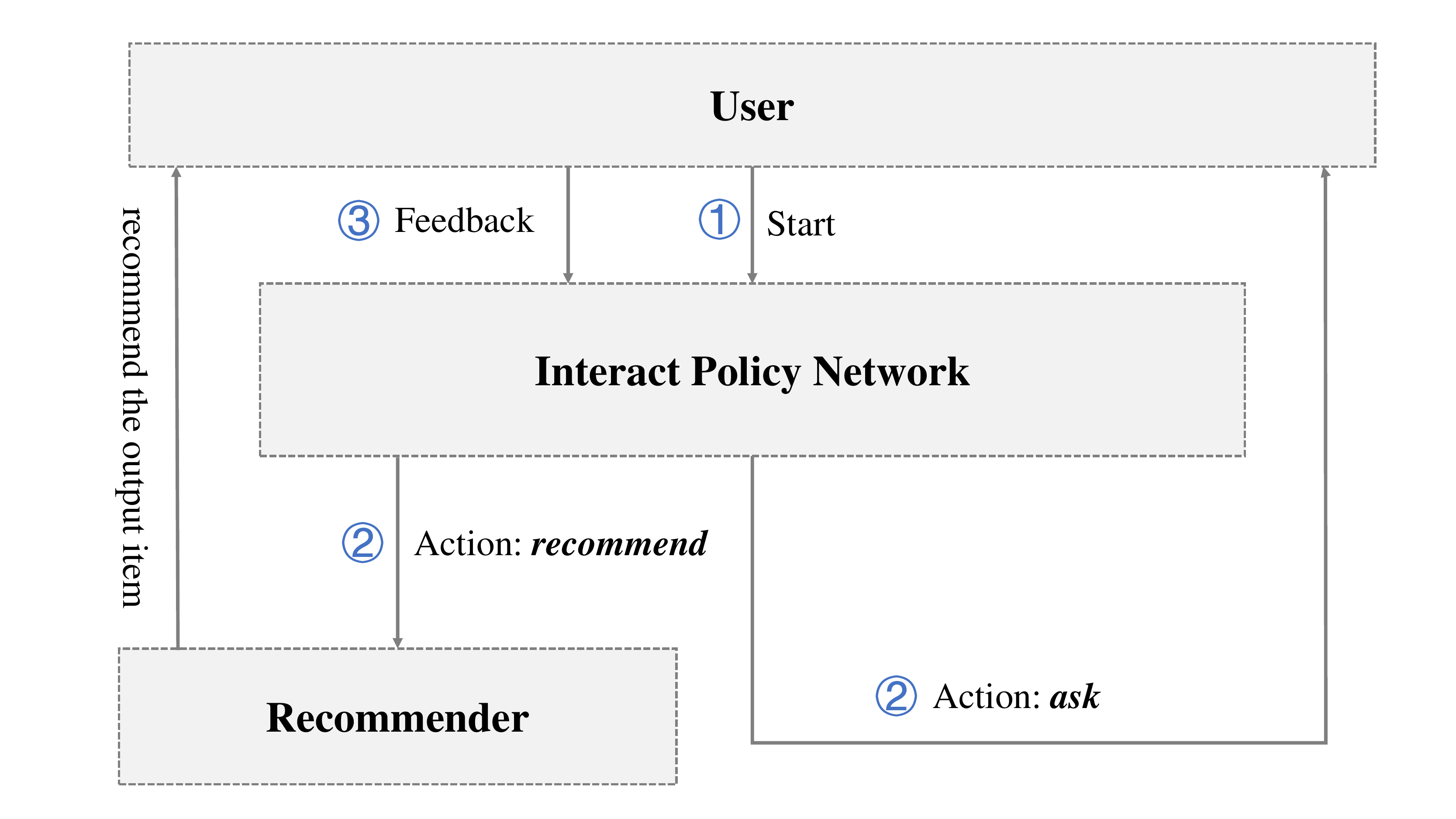}
\caption{The basic framework and workflow of multi-round CRS.}
\label{fig:CRSframework}
\end{figure}

As shown in Figure \ref{fig:CRSframework}, the user is the interaction object of CRS, and CRS mainly consists of a recommender module and an interactive policy module: the recommender module is responsible for generating a recommendation list based on the user preference information; the interactive policy module is responsible for deciding whether to ask users for more preference information or output recommended items at current conversational turn. As shown in Figure \ref{fig:CRSframework}, a conversational recommendation session starts when the user enters the system, and then CRS chooses the $recommend$ action or the $ask$ action according to the current understanding of the current user’s preferences. If the $recommend$ action is selected, the system will output the recommended items ranked to the top by the recommender to the user, and if the $ask$ action is selected, the system will output the asked item attributes selected by the interactive policy to the user. After that, back to the user side, the user gives binary feedback (accept or reject) on the recommended items or the asked item attributes. The whole interactive session ends when a successful recommendation is achieved or the set maximum round $T$ is reached.

\subsubsection{User Simulator}
User simulator is used to replace real users to interact with the system, making CRS tasks more suitable for offline academic research. We use the same user simulator design as \cite{sun2018conversational,lei2020estimation}.

\subsubsection{Recommender}
In the existing research work of CRS, researchers tend to choose the simple but effective recommendation model, Factorization Machine(FM), as the recommender of CRS. Because of its effectiveness and agility, FM can meet the recommendation requirements of CRS without increasing the complexity of CRS algorithm. Given user $u$ and his/her preferred attributes $P_u$ in the conversation, FM predict how likely $u$ will like current item $i$ and current attribute $p$ in the conversation session as:
\begin{equation}
\tilde{f}(i\mid u,P_u) = \boldsymbol{u}^T\boldsymbol{i}+\sum_{{p_k}\in{P_u}}\boldsymbol{i}^T\boldsymbol{p_k}
\end{equation}
\begin{equation}
\tilde{f}(p\mid u,P_u) = \boldsymbol{u}^T\boldsymbol{p}+\sum_{{p_k}\in{P_u}}\boldsymbol{p}^T\boldsymbol{p_k}
\end{equation}
The FM is trained by optimizing the pairwise Bayesian Personalized Ranking (BPR)\cite{2012RendleBPR} objective which assumes that the ground truth sample (item or attribute) should be ranked higher than other samples, and the training losses of the FM model are as follows:

\begin{sequation}
\begin{split}
{Loss}_{item} =\sum_{{(u,i^+,i^-)}\in{O_I}}&
-\ln\sigma (\tilde{f}(i^+\mid u,P_u)-\tilde{f}(i^-\mid u,P_u))\\
& + \lambda_{\Theta_{FM}} \|{\Theta_{FM}}\|^2
\end{split}
\end{sequation}

\begin{sequation}
\begin{split}
{Loss}_{attri} =\sum_{{(u,p^+,p^-)}\in{O_A}}& -\ln\sigma (\tilde{f}(p^+\mid u,P_u)-\tilde{f}(p^-\mid u,P_u))\\
& + \lambda_{\Theta_{FM}} \|{\Theta_{FM}}\|^2
\end{split}
\end{sequation}

\subsubsection{Interact Policy Network}
We use a two-layer feed forward neural network as the interact policy network and use the standard Deep Q-learning\cite{2019VolodymyrHuman} for optimization.
The policy network takes the current session state $state_his$ as input and outputs the values $Q(s, a)$ for the two actions, $a_{ask}$ and $a_{rec}$. CRS will choose the action with higher estimated reward $reward$, which is composed of $reward_{rec\_suc}$, $reward_{rec\_fail}$, $reward_{ask\_suc}$, $reward_{ask\_fail}$, $reward_{reach\_max\_turn}$. The value setting of reward function is shown in section \ref{sec:section 5.2.3}.

\section{Methodology}

\subsection{Overview}
\begin{figure*}[htb]
\centering
\includegraphics[width=7in]{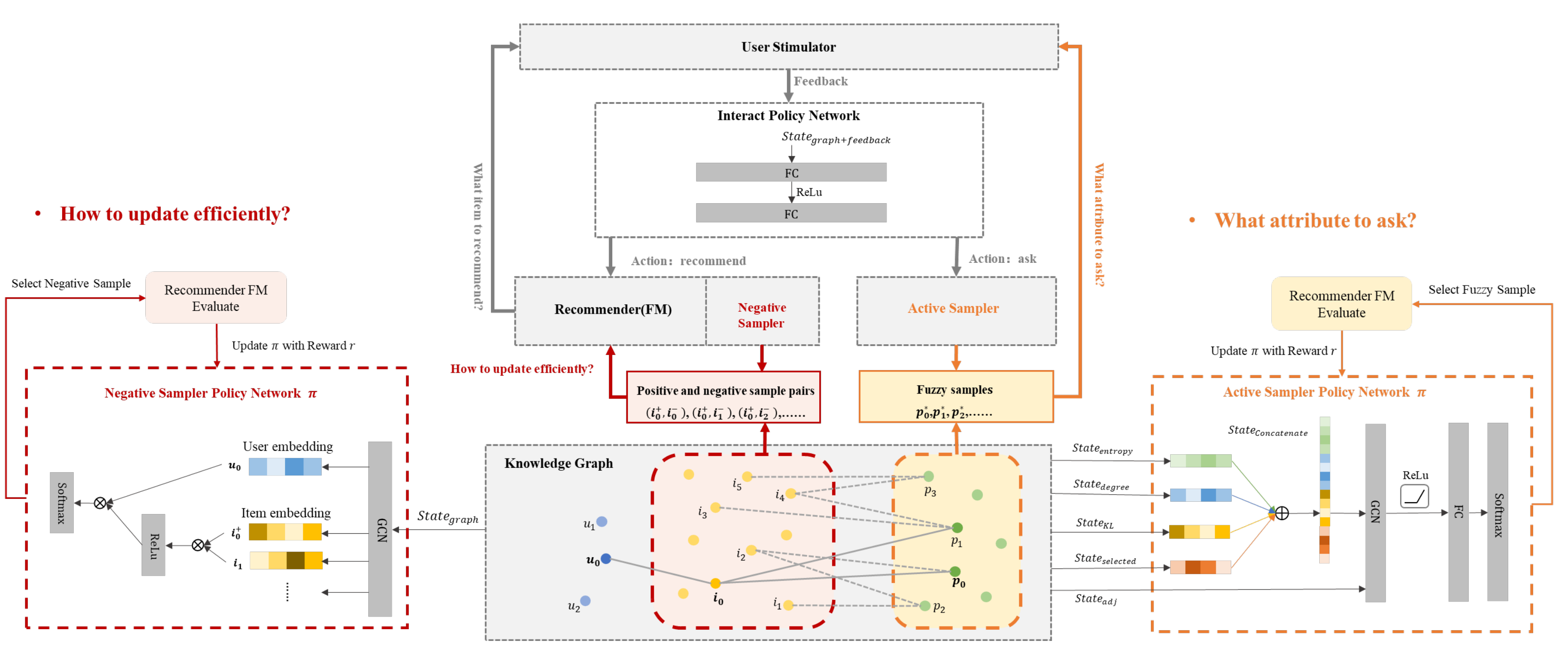}
\caption{KGenSam-based CRS framework overview. The key design is two sampler modules: (1) Negative Sampler outputs high-quality negative item samples and constructs positive and negative sample pairs to the effective update of the recommender at each turn of conversation; (2) Active Sampler outputs fuzzy item attribute samples with high uncertainty as the user preference information to be asked if CRS takes the $ask$ action.}
\label{fig:framework}
\end{figure*}

Figure \ref{fig:framework} illustrates the architecture of the proposed KGenSam, a general solution framework for knowledge-enhanced conversational recommendation. The middle subgraph in Figure \ref{fig:framework} shows the main body of a typical CRS. The workflow of the KGenSam CRS framework is as follows:

(1) Firstly, we construct the KG of user interaction data, and integrate the external KG to enrich the context information of interaction environment, and use knowledge to assist CRS interaction agent to make the decisions of actions.

(2) On the basis of the KG environment, we use the KG sampling methods\cite{2020HuGPA,2020WangKGPolicy} to enhance the knowledge and design two samplers based on reinforcement learning for CRS. Two samplers sample fuzzy samples and negative samples in KG respectively. Specifically, first, the \textbf{Active Sampler} focuses on the item attribute sample nodes representing user preferences in the KG, outputs fuzzy samples with large uncertainty through active reinforcement learning, and uses fuzzy samples for $ask$ action of CRS, so as to effectively improve the confidence of the model and improve the interaction efficiency of the system; second, the \textbf{Negative Sampler} focuses on the item sample nodes in the KG, outputs high-quality negative samples through reinforcement learning, supplements sparse online user data with negative samples, and updates the recommender with positive and negative sample pairs, so as to improve the update efficiency of the recommendation model. The two samplers work together to achieve the efficient acquisition of user preferences and reduce the interaction rounds, which reduce the user interaction burden.

The details of these two samplers modules will be discussed in section \ref{sec:section 4.2} and section \ref{sec:section 4.3} respectively.

(3) Finally, with the help of two samplers, the interactive policy network module learns the conversational strategy based on the enhanced knowledge, outputs the action ($ask$ or $recommend$) to be taken by the system, which guides the conversational direction, and realizes the acquisition of online user preferences and the successful recommendation of items that fit the user's current preferences at the last turn.

\subsection{Active Sampler via Active Reinforced Learning on KG}
\label{sec:section 4.2}

\begin{figure}[htb]
\centering
\includegraphics[width=3.5in]{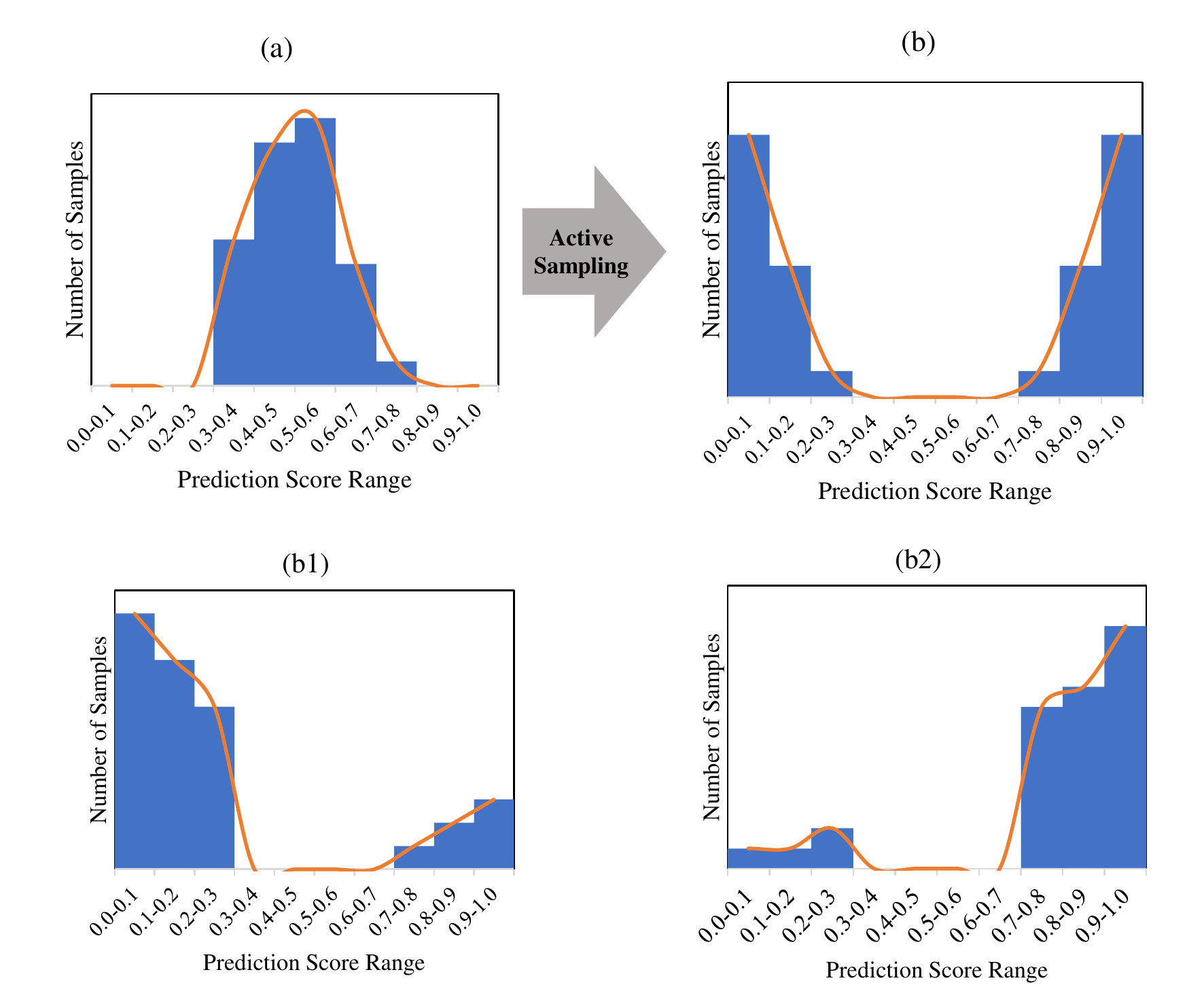}
\caption{Illustrations of the item attribute prediction score distribution of user preference output by the recommender. (a) When the recommender is not fully trained, the recommender's score on the item attribute is mostly 0.5, meaning that the user's preference for the attribute is uncertain. (b) When the recommender is fully trained, the prediction scores of item attributes are mostly close to 1.0 or 0.0, and the distribution curve of the sample is similar to $U$, and different users have different $U$, such as (b1) or (b2).}
\label{fig:ALdemopic}
\end{figure}
As discussed before, one of the problems in CRS research works is that "What attributes to ask ?", of which the specific research content is to help the recommender fit the current user's preferences efficiently. Corresponding to tackling this problem, the task of the Active Sampler is to choose fuzzy samples of item attributes to ask so as to shorten the conversation, as shown in Figure \ref{fig:ALdemopic}. In order to define the active learning process more easily, we regard the recommender as a two-classifier. For each sample, the recommender predicts whether belongs to user preferences, and the output result is between 0 and 1. Further, we cast active sampling as a Markov Decision Process (MDP) and design a policy network to learn the optimal $ask$ strategy.

\subsubsection{MDP of Active Sampling in KG}

In KG environment, there are 33 attributes in LastFM, 29 first-level attributes and 590 second-level attributes in Yelp. Considering that there are not many item attribute nodes, we put all attribute nodes that are not labeled by users into the candidate $\boldsymbol{Pool_{p^*}}$ (it should be noted here that it may be a better method to collect the attribute nodes with one or two hops of the current user node as the candidate pool for the dataset with too many attributes, which is worth trying in the future). Intuitively, given the condition of the current $state$ in KG, the Active Sampler takes an action by selecting the next fuzzy attribute node to $ask$. It is then rewarded by the performance gain of the FM trained with the updated set of labeled nodes. Formally, the key elements of MDP (MDP = \{$S, A, P, R$\}) are defined as follows.

\textbf{State.} The $\boldsymbol{state}$ vector is a concatenation of four component vectors that encode environmental information from different perspectives:
\begin{equation}
\begin{split}
    state=&{state}_{entropy}\oplus{state}_{degree}\oplus{state}_{KL}\\
    &\oplus{state}_{selected}\oplus{state}_{adj}
\end{split}
\end{equation}

\begin{itemize}
    \item $\bm{{State}_{entropy}.}$ Information theory often uses entropy to measure the degree of chaos and uncertainty of a system. The larger the entropy is, the more chaotic and uncertain the information of the system is. On the contrary, the smaller the entropy is, the more stable and deterministic the system is. In this binary classification scenario, the entropy of candidate samples is an important basis for the Active Sampler to sample.  Therefore, we calculate the entropy of the candidate attribute sample nodes to calculate the uncertainty of current environmental information, which is defined as follows :
    \begin{equation}
    -(\hat{y}_{\boldsymbol{Pool_{p^*}}}\log{\hat{y}_{\boldsymbol{Pool}}} +(1 - \hat{y}_{\boldsymbol{Pool_{p^*}}})\log{(1 - \hat{y}_{\boldsymbol{Pool_{p^*}}})})
    \end{equation}
    
    \item $\bm{{State}_{degree}.}$When calibrating an important node in the network, the relevant graph node centrality dimension is often used to define and measure the importance of the node\cite{2006KissEvaluatingCentrality}. Consequently, we use degree centrality, the most common graph node centrality dimension, to measure the importance of attribute sample nodes in KG. The definition is as follows :
    \begin{equation}
    \min{(degree(\boldsymbol{Pool_{p^*}})/\gamma,1)}
    \end{equation}
    where the hyperparameter $\gamma$ (set to 20) scale node degree and clip ${State}_{degree}$ to 1.
    
    \item $\bm{{State}_{KL}.}$ Kullback–Leibler (KL) divergence can measure the "distance" (asymmetric) between two probabilities, so KL divergence can be used to calculate those data samples with large deviation\cite{1990HurvichKL,2005SmithKL}. We compute the average KL divergence of the sample predicted label distribution between the attribute sample node and its neighbors , which can help the Active Sampler better identify potential clusters and decision boundaries in KG. The ${State}_{KL}$ measures local similarity as:
    \begin{equation}
    \begin{split}
        \frac{1}{\#neighbors}&
        (D_{KL}(\hat{y}_{\boldsymbol{Pool_{p^*}}}\|\hat{y}_{\boldsymbol{neighbors}})+\\
        & D_{KL}(\hat{y}_{\boldsymbol{neighbors}}\|\hat{y}_{\boldsymbol{Pool_{p^*}}}))
    \end{split}
    \end{equation}
    
    \item $\bm{{State}_{selected}.}$ For $p\in \boldsymbol{Pool_{p^*}}$, if  $p$ has been asked and responded, $\boldsymbol{{state}_{selected}}(p) = 1$; if $p$ has not been chosen yet, $\boldsymbol{{state}_{selected}}(p) = 0$.
    
    \item $\bm{{State}_{adj}.}$ In addition to the content information of the candidate attribute nodes, we also take the structural information of the current subgraph in KG as part of the $\boldsymbol{state}$. We construct the adjacency list dictionary of $\boldsymbol{Pool_{p^*}}$ as $\boldsymbol{{state}_{adj}}$ to take advantage of the KG structure.
    
\end{itemize}

\textbf{Action. }The space of actions is $\boldsymbol{Pool_{p^*}}$, and Active Sampler chooses an action $a^t$ from $\boldsymbol{Pool_{p^*}}$ based on the output results of active policy network, $\pi{(\cdot |state^{t-1})}$. As the state is changing during the interaction, the action space $\boldsymbol{Pool_{p^*}}$ is dynamic.

\textbf{Reward. } The $reward$ measures the quality of the $ask$ attribute sample at each turn. However, there is no ground truth in the selection of the best fuzzy samples in a active learning model, $reward$ rely on the feedback from the recommender. Consequently, we use the increase of the FM model's performance score from $t$ to $t+1$ turn to define the soft reward function and design the $reward$ as:
\begin{equation}
    reward^t=Gain\_Score(\hat{y}_{valid}^{t+1},\hat{y}_{valid}^{t})
\end{equation}
where $\hat{y}_{valid}^{t}$ is the evaluation score of the FM model at turn $t$ and $Gain\_Score$ calculates the increase of the score, where we use AUC as the evaluation metric.

\textbf{State Transition Dynamic. }Given an action $a^t$ at state $state^t$ , the transition to the next state $state^{t+1}$ is determined as:
\begin{equation}
    P(state^{t+1}=(u,a^t)|state^t,a^t)=1
\end{equation}

\textbf{Objective Function. }Towards learning a stochastic policy $\pi$, we maximize the expected cumulative discounted reward. The objective function with respect to the Active Sampler parameters $\Theta_{AS}$ is as follows:
\begin{equation}
    \max_{\Theta_{AS}} \sum_{{(u,p^+,p^-)}\in{O_A}} \mathbbm{E}_{\pi}[\sum_{t=1}^T \lambda_{AS}^{t-1}reward^t]
\end{equation}
where $\lambda_{AS}$ is the decay factor. 

\subsubsection{Active Policy Network}

 As the right subgraph in Figure \ref{fig:framework} shows, to effectively propagate useful information over the KG and thereby better measure node informativeness, we first parameterize the policy network $\pi$ as a one-layer Graph Convolutional Network (GCN)\cite{2016gcn}. Then We use ReLU as the activation function $\sigma$. On top of the GCN, we apply a linear layer to  output the final output embedding and get the score for each node. These scores are then normalized by a softmax layer to generate a probability distribution over all candidate attributes $p$ in $\boldsymbol{Pool_{p^*}}$.

By maximizing its long-term returns with policy gradient\cite{1999SuttonPolicyGradient}, active policy network can effectively learn to optimize the long-term performance of the recommender in an end-to-end fashion.

\subsection{Negative Sampler via  Reinforcement Learning on KG}
\label{sec:section 4.3}

After solving the problem that CRS asks what attributes to make each opportunity of $ask$ the most efficient, the next problem naturally arises, that is, how can the feedback signal of users for fuzzy samples be accurately learned by the recommender ?  In previous CRS work, all unobserved interactions are regarded as negative samples, so it is easy to produce low-quality negative samples, and it is difficult to update the model parameters of the recommender. Considering that CRS has very few session turns, the update of recommender need to be accurate and rapid.

In recommendation systems, users often directly ignore the uninterested items, which will not produce users' explicit negative feedback behavior and leads to the common One-Class Problem of recommender training data. To solve the negative sample missing problem caused by One-Class data, traditional recommendation systems often construct effective positive and negative sample pairs by negative sampling to learn user preferences. Drawing on the same idea, we design a Negative Sampler to help the recommender learn preference information as efficiently and timely as the Active Sampler's acquisition of user feedback information. The collected high-quality negative samples supplement the rare online negative feedback and user history positive samples. Similar to the active sampling, we cast negative sampling as a MDP and design a policy network to learn the features of high quality negative samples in conversational recommendation.

\subsubsection{MDP of Negative Sampling in KG}
In KG environment, different from the Active Sampler, the Negative Sampler needs to focus on the item nodes, so the candidate set to been sampled is very large. Since the Negative Sampler requires labor-intensive negative discovering and is memory-consuming and time-consuming to distill useful signals in large-scale KG. Thus, we design the Negative Sampler to navigate from a positive item over the KG structure like  $i^+\to p \to i$ and yield two-hops neighbors as candidate negative items . Conditioned on the navigated local subgraph $state$, the Negative Sampler takes an action by selecting a batch of possible negative samples from this two-hops candidate items and construct training pairs with ground-truth positive items. Similar to the reward design of Active Sampler, the reward of Negative Sampler is the performance score of the FM trained with the constructed positive and negative pairs. Formally, the key elements of MDP (MDP = \{$S, A, P, R$\}) are defined as follows.

\textbf{State. }Given user $u$, the $state^t$ at turn $t$ is defined as subgraph paths $(u\to \cdots \to i^t)$, where $i^t$ is the node the Negative Sampler visits currently. 

\textbf{Action. }The space of actions is $\boldsymbol{Pool_{i^-}}$, which is the candidate sets of two hop neighbors.

\textbf{Reward. }The design goal of reward is to measure the quality of negative samples. Similar to the concept of hard negative in the field of Image Detection, we need to find the discriminative negative samples which are easily misjudged by the recommender in the CRS task. Inspired by the negative sampling model (KGPolicy) proposed by Wang et al.\cite{2020WangKGPolicy}, we apply the hypothesis and definition of "Real Negative Sample" to the negative sampling task scenario of CRS. The definition of high-quality negative sample has two characteristics: it is similar to the user representation\cite{park2018Adversarial,wang2017IRGAN}; it is close to the real positive sample. The negative samples with these two characteristics can represent the real negative samples and provide a larger gradient for the parameter update of the recommender. From this point of view, we define the reward of negative sampler as follows.
\begin{equation}
\begin{split}
    reward^t=(\boldsymbol{{Pool_{i^-}^t}})\mathsf{T} \cdot \boldsymbol{u}+(\boldsymbol{Pool_{i^-}^t})\mathsf{T} \cdot \boldsymbol{i^+}
\end{split}
\end{equation}

\textbf{State Transition Dynamic. }Given an action $a^t$ at state $state^t$ , the transition to the next state $state^{t+1}$ is determined as:
\begin{equation}
    P(state^{t+1}=(u,a^t)|state^t,a^t)=1
\end{equation}

\textbf{Objective Function. }To maximize the sum of expected rewards obtained from following policy $\pi$ over the training graphs, the objective function with respect to the negative policy network parameters $\Theta_{NS}$ is:
\begin{equation}
    \max_{\Theta_{NS}} \sum_{{(u,i^+,i^-)}\in{O_I}} \mathbbm{E}_{\pi}[\sum_{t=1}^T \lambda_{NS}^{t-1}reward^t]
\end{equation}
where $\lambda_{NS}$ is the decay factor.

\subsubsection{Negative Policy Network}

As shown in the left side of Figure \ref{fig:framework}, the Negative Sampler first obtains the initial representation of the nodes through one layer of GCN, which captures the knowledge-aware negative signals, and then output a batch of negative item nodes with high weight through the attention model constructed by two-layer activation function.


\section{Experiments and Analysis}
\subsection{Experimental Setup}
\subsubsection{Datasets}
As summarized in Table \ref{table:datasets}, we use two publicly available datasets: LastFM and Yelp. Each dataset is composed of two components, the user-item interactions and KG derived from external data sources, Freebase for LastFM and local business information network for Yelp. The cut ratio of training, validation and testing sets is 7: 1: 2. The yelp dataset we use here is slightly different from EAR\cite{lei2020estimation} and CPR, using 590 second-level attributes instead of 29 first-level attributes. Thus, for real application scenarios, LastFM represents the dataset with fewer attributes, Yelp represents the dataset with more attributes.

\begin{table}[!t]
\caption{Dataset Statictics of LastFM and Yelp}
\label{table:datasets}
\centering
\begin{tabular}{c|c|c|c}
\toprule[1pt]
                   &      &\textbf{LastFM}  &\textbf{Yelp} \\ \midrule[0.5pt]
\multirow{4}*{User-Item Interaction}   
                   ~& \#Users   & 1,801   &27,675\\
                   ~& \#Items   & 7,432	&70,311\\  
                   ~& \#attributes  & \textbf{33}	&\textbf{590}\\ 
                   ~& \#Interactions   & 76,693  &1,368,606\\ 
                   \hline
\multirow{2}*{Knowledge Graph}  
                   ~& \#Entities  & 9,266    &98,576\\
                   ~& \#Triplets  & 138,217    &2,533,827\\  
\bottomrule[1pt]
\end{tabular}
\end{table}

\subsubsection{Parameter Setting}
The maximum turn $T$ is set as 15. 
All hyperparameters for offline training of FM remains the same as EAR\cite{lei2020estimation}. The detailed rewards and the hyperparameters of interact policy net remain the same as CPR\cite{lei2020interactive}. All the hyperparameters related samplers training are tuned according to the validation set.

\subsubsection{Training Steps}
The training process is made up of three stages.

\textbf{Stage 1:} Pretraining of recommender for scoring items and attributes. Strictly following\cite{lei2020estimation}, we use the training set of $O_I$ to optimize FM offline. We refer the readers to the original paper \cite{lei2020estimation} for more Information. 

\textbf{Stage 2:} Pretraining of two samplers. The pretrained FM is used to provide reward scores for two sampling networks. The detailed pretraining process of samplers are shown in Algorithm 1 and 2 respectively.

\textbf{Stage 3:} Online training of interact network. We use a user simulator to interact with the user to train the policy network. The interaction algorithm flow of CRS based on our KGenSam is shown in Algorithm 3.

\begin{algorithm}[ht]
	\caption{Pretrain the Active Sampler}
	\begin{algorithmic}
	    \REQUIRE $O_A$, FM
		\ENSURE Active Sampler Policy $\pi$ 
		\FOR{$(u,p^+,p^-)\in O_A$}
		    \STATE $Pool_{p^*} \leftarrow A$ 
    		\FOR{$t=1$ to $T$}
    			\STATE select a batch ranked $p^*_t$ from $Pool_{p^*}$
    			\IF{$p^*_t$ in $A^+_u$}
    			    \STATE $p^+_u \leftarrow p^*_t$ 
    			    \STATE update FM based on $p^+_u$
    			\ELSE
    			    \STATE $p^-_u \leftarrow p^*_t$ 
    			\STATE update FM based on $p^-_u$
    			\ENDIF
    			\STATE remove $p^*_{t}$ from $Pool_{p^*}$
    		\ENDFOR
		    \STATE evaluate FM on the validation set to get the reward signal $\sum_{t=1}^T reward^t$
		    \STATE update $\pi$ with policy gradient
		\ENDFOR
		
	\end{algorithmic}
\end{algorithm}

\begin{algorithm}[ht]
	\caption{Pretrain the Negative Sampler}
	\begin{algorithmic}
	    \REQUIRE $O_I$, FM
		\ENSURE Negative Sampler Policy $\pi$ 
		\FOR{$(u,i^+,i^-)\in O_I$}
		    \STATE $ Batch_{i^-} \leftarrow \{\}, Pool_{i^-} \leftarrow \{\}, i_s \leftarrow i^+ $ 
    		\FOR{$s=1$ to $steps$}
    			\STATE get neighbors $Neighbors_{i_s}$ of $i_s$
    			\STATE traverse $Neighbors_{i_s}$ as $i_s$
    			\FOR{each $i_{neighbor} \in Neighbors_{i_s}$}
    			    \IF{$i_{neighbor} \notin I^+$}
    			        \STATE add $i_{neighbor}$ to $Pool_{i^-}$
    			    \ENDIF
    			\ENDFOR
			\STATE select a batch ranked $Batch_{i^-}$ from $Pool_{i^-}$
			\STATE update FM based on $Batch_{i^-}$ and $i^+$
		    \STATE evaluate FM on the validation set to get the reward signal $\sum_{t=1}^B reward^t$
		    \STATE update $\pi$ with policy gradient
    		\ENDFOR
		\ENDFOR
		
	\end{algorithmic}
\end{algorithm}

\begin{algorithm}[ht]
	\caption{KGenSam-based CRS Algorithm}
	\begin{algorithmic}
	    \REQUIRE $O_I$, $O_A$, FM, Active Sampler, Negative Sampler
		\ENSURE Conversation Recommendation Policy $\pi$ 
		\FOR{$(u,i^+,i^-)\in O_I, (u,p^+,p^-)\in O_A $}
		    \STATE $P^+_u \leftarrow \{p^+_0\}, P^-_u \leftarrow \{\}, Pool_{p^*} \leftarrow A$
		    \STATE $Pool_{i^-} \leftarrow Neighbors_{i^+}-I^+ $ 
    		\FOR{$t=1$ to $T$}
    		    \STATE select an action $a_t$ from $\{ask,rec\}$
    		    \STATE Negative Sampler selects negative batch $Batch_{i^-}$ 
    			\IF{$a_t = ask$} 
        			\STATE remove $p^+_{t-1}$ from $Pool_{p^*}$
        			\STATE Active Sampler selects a $p^*_t$ from $Pool_{p^*}$
        			\IF{user $u$ accepts $p^*_t$}
        			    \STATE add $p^*_t$ to $P^+_u$
        			\ELSE
        			    \STATE add $p^*_t$ to $P^-_u$
        			\ENDIF
        			
    		    \ELSE 
        			\STATE FM outputs $i^+_t$ to recommend
        			\IF{user $u$ accept $i^+_t$}
        			    \STATE achieve successful recommendation and quit!
        			\ELSE
        			    \STATE add $i^+_t$ to $Batch_{i^-}$
        			\ENDIF
        			
    		    \ENDIF 
    		    \STATE update FM based on $P^+_u$, $P^-_u$ and $Batch_{i^-}$ 
    		\ENDFOR
		
		    \STATE reach the maximum turns and quit! 
		\ENDFOR
		
	\end{algorithmic}
\end{algorithm}

\subsubsection{Comparison Methods}
We compare our method against the following baseline methods:
\begin{itemize}
    \item Abs Greedy\cite{Christakopoulou2016TCRS}.This method serves as a baseline where the model only recommends items at each turn and updates itself until it finally makes successful recommendation.
    \item Max Entropy. This method is based on the maximum entropy rule. It always chooses an attribute with the maximum entropy within the current candidate item set\cite{Dhingra2016End} to ask. Details can be found at \cite{lei2020estimation}.
    \item CRM\cite{sun2018conversational}. This is a CRS model which is originally designed for single-round recommendation. CRM records user’s preference into a belief tracker, and uses reinforcement learning to learn the policy of interaction. We follow \cite{lei2020estimation} to adapt it to the multi-round scenario.
    \item EAR \cite{lei2020estimation}. This is the state-of-the-art method which proposed a three stage solution, Estimation Action Reflection. 
    \item SCPR \cite{lei2020interactive}. This is the simple implementation model of the state-of-the-art work, CPR, which introduces KG in CRS for the first time.
    \item FPAN \cite{xv2021FPAN}. This is the latest CRS research work. It pays attention to the correction effect of user feedback on recommendation and adjusts the node representation of KG through the Gate Mechanism, which significantly improves the CRS recommendation results hence being the most comparable model.
\end{itemize}

\subsubsection{Evaluation Metrics}
The evaluation follows \cite{lei2020estimation}. We use the success rate at the 15th turn of conversation (SR@15) to measure the cumulative ratio of successful recommendation by turn 15 . We also use and average average turns (AT) of conversation length when the interaction process ends to record the average number of turns for a successful recommendation. Therefore, the higher SR@15 and the lower AT indicate a higher performance and overall higher efficiency.

\textbf{Success Rate@15.}
\begin{equation}
    SR@15=\frac{1}{\#users \times T }\sum_{u\in U}\sum_{t=1}^T \theta_{hit}
\end{equation}

where $\theta_{hit} = 1$, if the instinctive feedback of the recommended item given by the simulator is acceptance.

\textbf{Average Turns.}
\begin{equation}
    AT=\frac{1}{\#users} \sum_{u\in U}{T_{ends}}
\end{equation}

where $T_{ends}$ is the total turns when the current conversation  session ends.

\subsection{Results and Discussion}

\subsubsection{Overall Performance Comparison}

\begin{table}
\caption{Success Rate@15 and Average Turns}
\label{table:compareall}
\centering
\begin{tabular}{c|c|c|c|c}
\hline
~ & \multicolumn{2}{c|}{\textbf{LastFM}} &\multicolumn{2}{c}{\textbf{Yelp}}\\
      \hline
    ~ &  \textbf{SR@15} &  \textbf{AT} &  \textbf{SR@15} &  \textbf{AT}\\
      \hline
    Abs Greedy &  0.222 &  13.48 &  0.189 &  13.43\\ 
    Max Entropy &  0.283 &  13.91 &  0.398 &  13.42\\
    CRM &  0.325 &  13.75 &  0.177 &  13.69\\ 
    EAR &  0.429 &  12.88 &  0.182 &  13.63\\ 
    SCPR &  0.465 &  12.86 &  0.489 &  12.62\\ 
    FPAN &  0.667 & 10.14 &  0.588 &  12.65\\
      \hline
    \textbf{KGenSam} &  \textbf{0.948} &  \textbf{6.43} &  \textbf{0.810} &  \textbf{7.72}\\
      \hline
    improve  & 42.0\%↑ & 36.6\%↑ & 37.8\%↑ & 38.8\%↑\\
     \hline
\end{tabular}
\end{table}

\begin{figure*}[htb]
\centering
\includegraphics[width=7in]{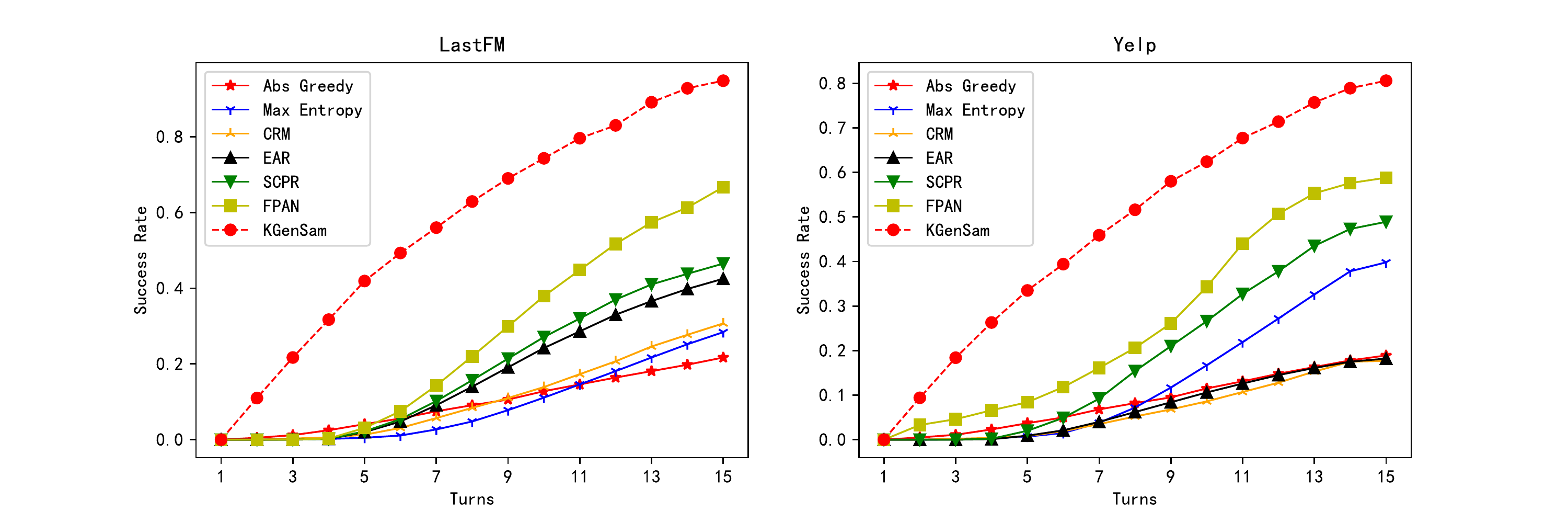}
\includegraphics[width=7in]{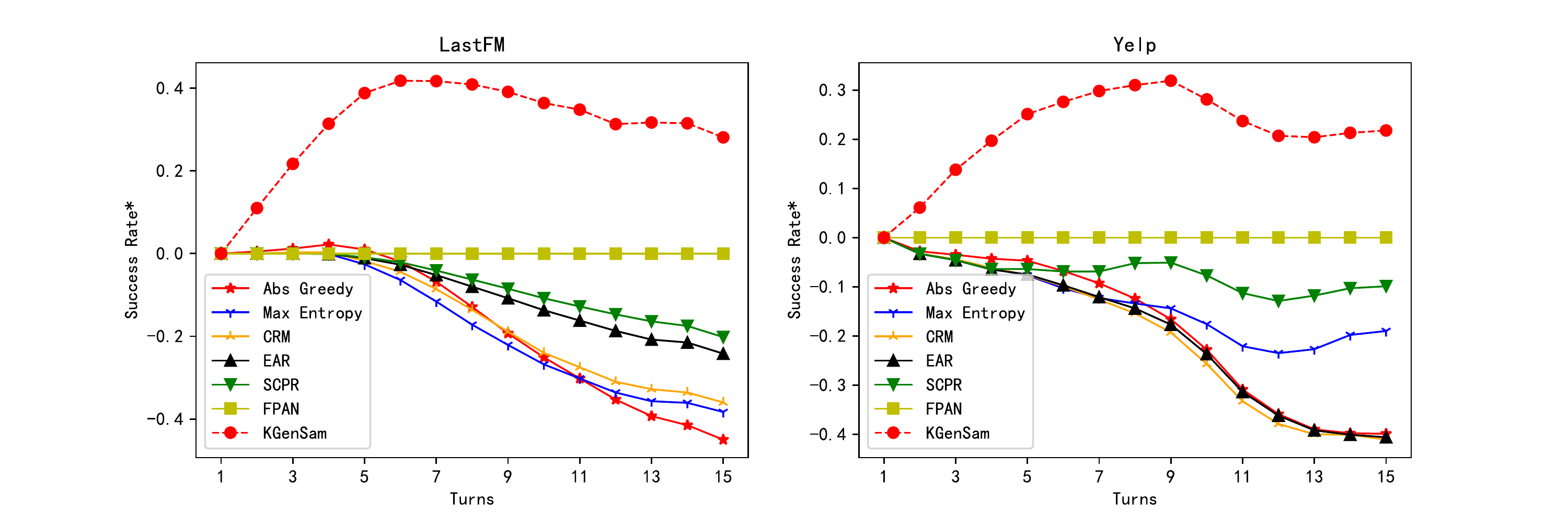}
\caption{SR of compared methods at different turns on LastFM and Yelp.}
\label{fig:compareall}
\end{figure*}

The results are summarized in Table \ref{table:compareall}. Overall, it can be seen that the proposed KGenSam architecture outperforms other baselines in both terms of SR@15 and AT, which validates our hypothesis that enhancing the information modeling of interactive environment is an effective strategy to improve the performance of CRS. 

By intuitively presenting the performance comparison in Figure \ref{fig:compareall}, we also have following discoveries:

\begin{itemize}
    \item For Yelp dataset which has large attribute space, the performance of all baseline methods is not satisfactory, CRM and EAR models can not even achieve the performance of Max Entropy method.
    \item From the improvement trend of success rate in the whole conversation, we can clearly find that the first six rounds of interaction of the baseline methods are in a "Chilling Period", during which the success rate of recommendation is very low, and after the sixth round, the success rate of CRS slowly increases. However, KGenSam does not have such a "Chilling Period", which can improve the success rate after each round of interaction, and the improvement of the early interaction is even greater than that of the later interaction, which is consistent with the trend of the accuracy improvement curve of Active Learning in the classification model annotation task. This consistency verifies our hypothesis that the idea of active learning can effectively integrate with CRS task.
\end{itemize}

\subsubsection{Evaluating Key Designs in KGenSam}

\begin{table*}[!ht]
\caption{Performance of removing Samplers from KGenSam}
\label{table:ablation}
\centering
\resizebox{\textwidth}{13mm}{
\begin{tabular}{c|c|c|c|c|c|c|c|c|c|c|c|c}
\hline
~ & \multicolumn{6}{c|}{\textbf{LastFM}} &\multicolumn{6}{c}{\textbf{Yelp}}\\
      \hline
    ~ &  \textbf{SR@3} &  \textbf{SR@6} &  \textbf{SR@9} &  \textbf{SR@12} &  \textbf{SR@15} &  \textbf{AT} &  \textbf{SR@3} &  \textbf{SR@6} &  \textbf{SR@9} &  \textbf{SR@12} &  \textbf{SR@15} &  \textbf{AT}\\
      \hline
    w/o Active Sampler &  0.000 &  0.085 &  0.324 &  0.561 &  0.7225 (23.8\%↓) &  9.18 &  
    0.000 &  0.036 &  0.143 &  0.247 &  0.309 (61.9\%↓) &  12.94\\ 
    w/o Negative Sampler &  0.157 &  0.355 &  0.491 &  0.602 &  0.673 (29.0\%↓) &  10.55 &  
    0.135 &  0.288 &  0.417 &  0.514 &  0.572 (29.4\%↓) &  10.23\\
    w/o Samplers &  0.000 &  0.051 &  0.204 &  0.353 &  0.455 (52.0\%↓) &  12.65 &  
    0.000 & 0.000 &  0.023 &  0.090 &  0.195 (75.9\%↓) &  12.59\\
      \hline
    \textbf{KGenSam} &  \textbf{0.217} &  \textbf{0.493} &  \textbf{0.690} &  \textbf{0.830} &  \textbf{0.948} &  \textbf{6.43} &  
    \textbf{0.184} &  \textbf{0.394} &  \textbf{0.580} &  \textbf{0.714} &  \textbf{0.810} &  \textbf{7.72}\\
      \hline
\end{tabular}}
\end{table*}

\begin{figure*}[htb]
\centering
\includegraphics[width=7in]{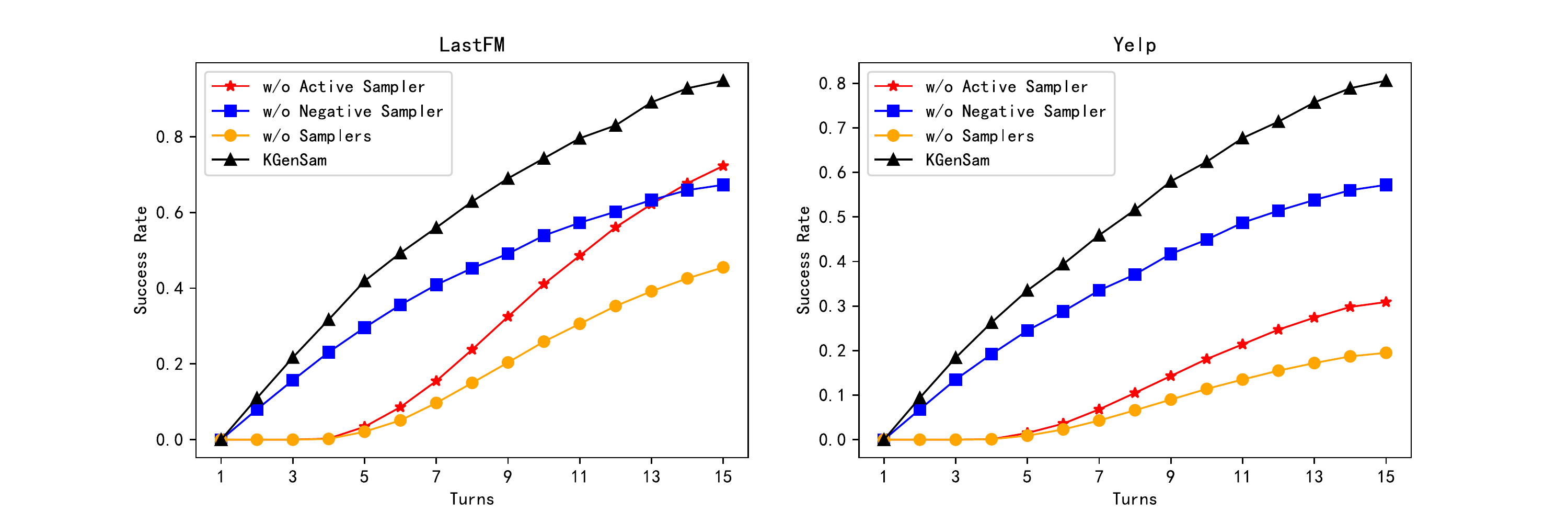}
\includegraphics[width=7in]{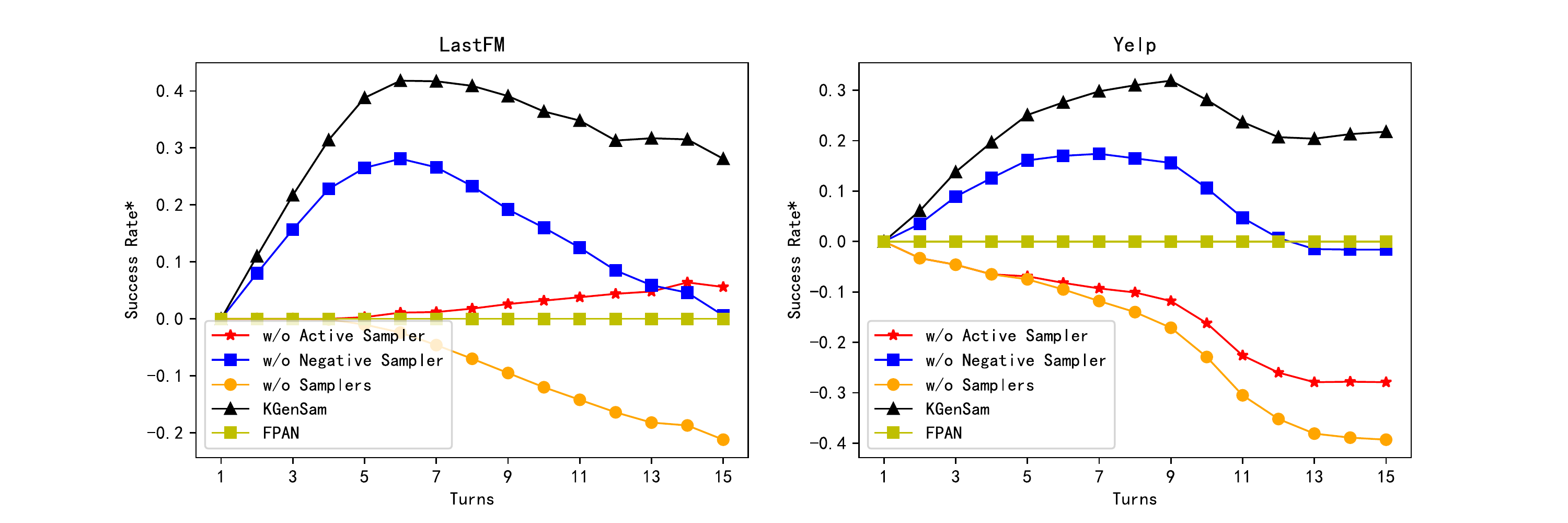}
\caption{Success rate of different ablation setting KGenSam on LastFM and Yelp.}
\label{fig:ablation}
\end{figure*}
The key designs of KGenSam are the two sampler modules. In order to evaluate whether the two samplers can meet our expectations, that is, to accurately select the asked attribute samples and efficiently update the recommender, we designed the following ablation experiments.  

\textbf{(1) w/o Active Sampler :} We select attribute samples with the maximum scores (calculated by propagating messages like CPR\cite{lei2020interactive}) to ask.  

\textbf{(2) w/o Negative Sampler :} We randomly select item samples without interaction as negative samples to update the recommender.  

\textbf{(3) w/o Samplers :} Neither of the two sampling modules works. We only use KG as the environment, and the selection methods of asked attributes and negative items are consistent with (1) and (2).

Table \ref{table:ablation} shows the success rate of every three turns and the average rounds of a session under the above three ablation experimental settings. It can be seen that when two samplers work at the same time, the performance of the complete KGenSam is the best. 

In addition, we have the following conclusions based on the success rate trend in Figure \ref{fig:ablation}:

\begin{itemize}
    \item The Active Sampler changes the shape of the success rate curve and makes it convex upward, indicating that the Active Sampler changes the magnitude of user preference information acquisition per turn. As indicated in Figure \ref{fig:ablation}, the Active Sampler can greatly improve the performance of CRS in the first six rounds of the dialogue, which means it can help CRS solve the problem of "Chilling Period" (the phenomenon mentioned in the previous section). Figure \ref{fig:ablation} illustrates that the fuzzy attribute samples output by the Active Sampler are indeed the most uncertain user preferences in the current system, which means the attribute knowledge enhanced by the Active Learning idea can help CRS ask the user preference information accurately.
    \item As shown in Figure \ref{fig:ablation}, from the blue line to the black line and from the orange line to the red line, we can see that the Negative Sampler makes the success rate curve move upward as a whole, which indicates that the Negative Sampler improves the learning speed of user preference information, and makes the success rate rise to a relatively high level earlier.
    \item From Figure \ref{fig:ablation}, we can also notice that KGenSam with Active and Negative Samplers removed separately behaves differently on the two datasets. For the LastFm with fewer attributes, the effect of removing the active sampler is less than that of removing the Negative Sampler; while for Yelp with more attributes, the opposite is true. Through the analysis of this result, we believe that the Active Sampler can play a greater role in the complex data environment because it focuses on the selection of attribute nodes. Thus, the Active Sampler has a greater impact on Yelp which has a large number of attributes.
    \item Table \ref{table:ablation} and Figure \ref{fig:ablation} verify that the Active Samplers are responsible for obtaining a large amount of user preference information, and the Negative Samplers are responsible for learning the acquired information. Two key designs achieve a high-performance CRS by jointly enhancing the environmental information in KG.
\end{itemize}

\subsubsection{Performance under Different Reward}
\label{sec:section 5.2.3}

\begin{table}
\caption{Reward Settings of Interact Policy Network in CRS}
\label{table:reward}
\centering
 \setlength{\tabcolsep}{1mm}{
\begin{tabular}{c|c|c|c|c}
    \hline
    &  $\boldsymbol{R_{CPR}}$ &  $\boldsymbol{R_{ask\_more}}$ &  $\boldsymbol{R_{rec\_more}}$ &  $\boldsymbol{R_{EAR}}$\\
     \hline
    $reward_{ask\_suc}$ &  0.01 &  0.1 &  0.01 &  0.01+0.1\\ 
    $reward_{ask\_fail}$ &  -0.1 &  -0.1 &  -0.1 &  0.01+0\\
    $reward_{rec\_suc}$ &  1 &  1 &  1 &  0.01+1\\ 
    $reward_{rec\_fail}$ &  -0.1 &  -1 &  -0.01 &  0.01+0\\ 
    $reward_{reach\_max\_turn}$ &  -0.3 &  -0.3 &  -0.3 &  -0.3\\
      \hline
\end{tabular}}
\end{table}

\begin{figure}[htb]
\centering
\includegraphics[width=3in]{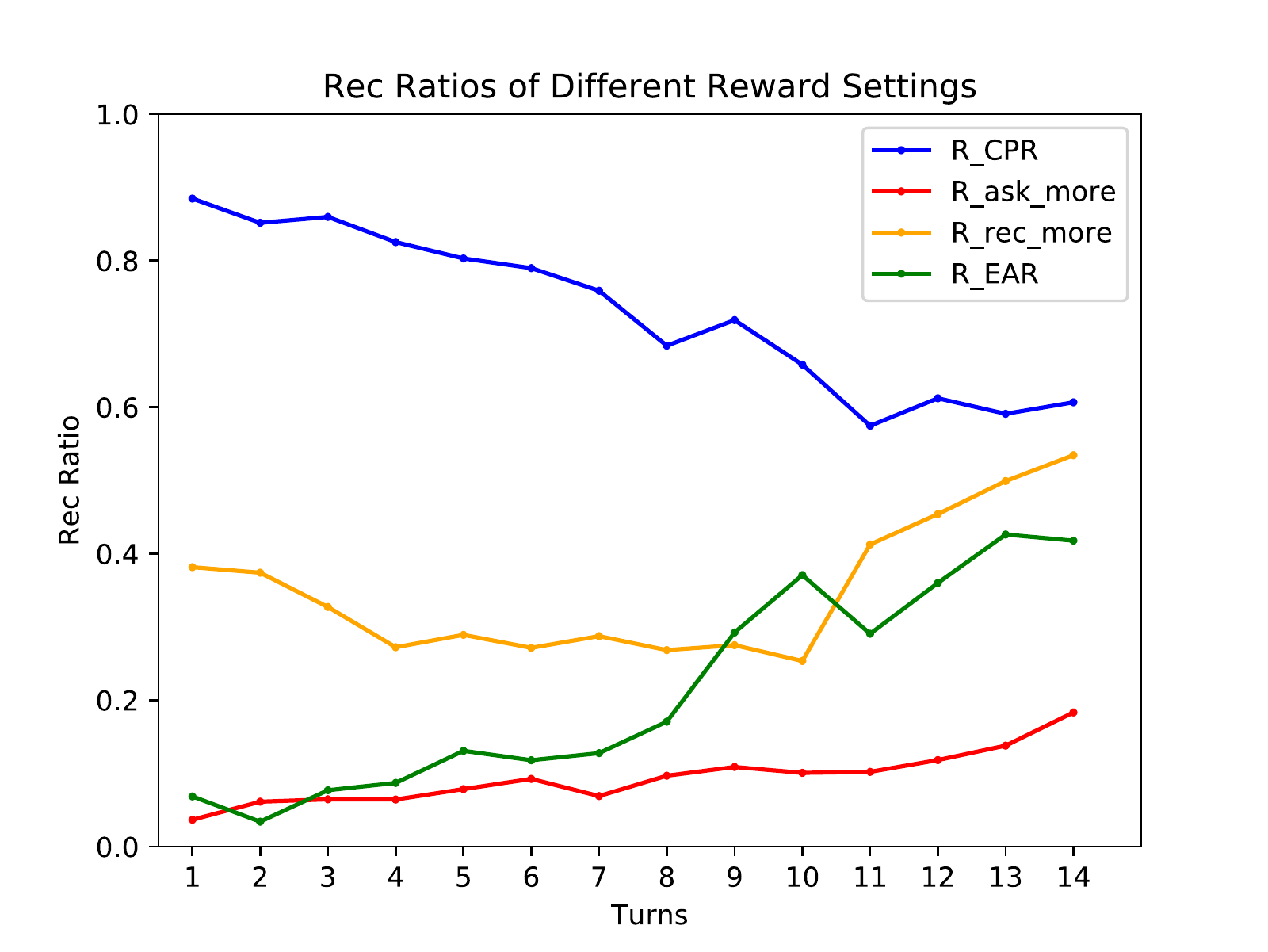}

\caption{Rec Ratios of KGenSam-based CRS trained with different reward settings in LastFM.}
\label{fig:recratio}
\end{figure}

In the process of experiment, we found a phenomenon worthy of attention, that is, the reward setting of interact policy network has a great impact on the final evaluation result of CRS, which may be due to the current evaluation metrics of CRS is not comprehensive enough. For example, the value settings of reward in the two representative CRS works, EAR and CPR, are totally different. We use the same reward setting of CPR which is also based on KG as the default reward setting in our KGenSam framework. In order to further verify the effect of reward setting in CRS, we design several kinds of reward settings as shown in Table \ref{table:reward}, in which $reward_{ask\_suc}$, $reward_{ask\_fail}$, $reward_{rec\_suc}$, $reward_{rec\_fail}$ and $reward_{reach\_max\_turn}$ represent "ask attribute succeed", "ask attribute fail", "recommend succeed", "recommend fail" and "reach maximum turn" respectively. $R_{CPR}$(default setting) and $R_{EAR}$ are the reward settings of CPR and EAR. In addition, we increase $reward_{ask\_suc}$ and decrease $reward_{rec\_fail}$ to set $R_{ask\_more}$ which can tend to $ask$, in the same way, we set $R_{rec\_more}$ to incline $recommend$. 

It can be seen from Figure \ref{fig:recratio} that reward setting can affect the probability of taking $recommend$ action. The reward setting of CPR makes the rec ratio of each round extremely high, and the final performance on both SR@T an AT is also the best, which verifies our previous conjecture. We boldly draw the conclusion that rec ratio is positively correlated with the two evaluation indicators, which needs to be verified by more research in the future.

\section{Conclusion}

In this paper, we provide a new idea for the research of CRS, that is to solve the problem of insufficient environmental information modeling in CRS from the perspective of knowledge enhancement. The experimental results show that this perspective is reliable and effective. We construct an Active Sampler to focus on the fuzzy attribute nodes in the graph to maximize the amount of information obtained by each $ask$ turn. At the same time, the Negative Sampler we constructed focus on the high-quality negative samples in KG, so as to maximize the learning speed of each recommender update. As far as we know, we are the first to define the idea of Active Learning into the CRS problem and achieve the experimental results overcoming "Chilling Period" problem. This idea of Active Learning to screen fuzzy samples in KG can also be applied to the cold start problem of KG based recommendation system. In addition, we also introduce hard negative Samples into CRS to achieve efficient CRS online update and promote the commercial online application of CRS.

In the future, we will try to introduce more evaluation metrics to solve the incomplete evaluation problem in CRS research. In addition, the current CRS research is based on single session (including single-round and multi-rounds). Our KGenSam method has achieved high evaluation results in the single session scenario where the conversation ends after only one successful recommendation, but we believe that single-session CRS is obviously not in line with the real recommendation scenario. Therefore, from the perspective of sequential recommendation modeling, we will study multi-session modeling of CRS 
to extend our KGenSam framework, which may improve the possibility of CRS being widely used in business scenarios.


%



\ifCLASSOPTIONcompsoc
  \section*{Acknowledgments}
\else
  \section*{Acknowledgment}
\fi

This work is supported by the National Key R\&D Program of China (2018AAA0100604), the Fundamental Research Funds for the Central Universities (2021RC217), the Beijing Natural Science Foundation (JQ20023), the National Natural Science Foundation of China (61632002, 61832004, 62036012, 61720106006).

\ifCLASSOPTIONcaptionsoff
  \newpage
\fi



\bibliographystyle{IEEEtran}
\bibliography{references}
%

%








\end{document}